\documentclass[12pt]{iopart}

\usepackage{iopams,amsfonts,amssymb,amsthm} 

\pdfoutput=1
\usepackage{etex}
\usepackage{cite}
\usepackage{latexsym}
\usepackage{rawfonts}
\usepackage{graphicx}
\usepackage[usenames,dvipsnames,svgnames]{xcolor}
\usepackage{bbm}
\usepackage{latexsym}
\usepackage{multirow}
\usepackage{rotating}
\usepackage{lscape}
\usepackage{graphicx} 
\usepackage{subcaption}
\usepackage{xcolor}
\usepackage{fancybox}
\usepackage{ifmtarg}
\usepackage{fancyhdr}
\usepackage{wrapfig}
\usepackage{hyperref}
\usepackage{tikz}
\usepackage{ulem}
\usepackage[flushleft]{threeparttable}

\input prepictex
\input pictex
\input postpictex

\def\2;{\;\;}

\def\eps{\epsilon}

\def\Ref#1{(\ref{#1})}

\def\C#1{{\mathcal #1}}

\def\Sfrac#1#2{\hbox{\large $\frac{#1}{#2}$}}

\def\LB{\left(}         \def\RB{\right)}

\def\LC{\left\{}       \def\RC{\right\}}

\def\vv{{\;\hbox{\Large $|$}\;}}

\def\tiny#1{\scalebox{0.75}{#1}}

\definecolor{blue}{rgb}{0,0.18,0.39}
\definecolor{RoyalBlue}{rgb}{0,0.2,0.7}

\definecolor{Maroon}{cmyk}{0,0.87,0.68,0.62}
\definecolor{Brown}{rgb}{0.7,0.3,0}
\definecolor{Navy}{rgb}{0.3,0.0,0.4}
\definecolor{Red}{cmyk}{0,1,1,0}
\definecolor{BrickRed}{cmyk}{0.16,0.89,0.61,0.02}
\definecolor{DarkRed}{cmyk}{0,1,1,0.5}
\definecolor{DarkBlue}{cmyk}{1,1,0,0.2}
\definecolor{DarkGreen}{cmyk}{1,0,1,0.4}
\definecolor{Green}{cmyk}{1,0,1,0}
\definecolor{DarkBrown}{cmyk}{0,0.81,1,0.6}
\definecolor{OrangeRed}{cmyk}{0,1,0.87,0}
\definecolor{RedOrange}{cmyk}{0,0.77,0.87,0}
\definecolor{Orange}{cmyk}{0,0.61,0.87,0}
\definecolor{Offwhite}{rgb}{.8,0.9,.8}
\definecolor{Offwhite2}{cmyk}{.04,.02,.01,0}
\definecolor{Tan}{rgb}{0.82,0.70,0.55}
\definecolor{Blue}{rgb}{0,0,1}
\definecolor{RoyalBlue}{rgb}{0.25,0.41,0.88}
\definecolor{Sepia}{rgb}{0.37,0.14,0.07}
\definecolor{myblue}{cmyk}{0.025,0.05,0,0}
\definecolor{Mahogany}{cmyk}{0.18,0.87,1,0.08}

\definecolor{green1}{cmyk}{0.25,0,0.76,0}
\definecolor{green2}{cmyk}{0.25,0,0.76,0.07}
\definecolor{green3}{cmyk}{0.25,0,0.76,0.20}
\definecolor{green4}{cmyk}{0.25,0,0.75,0.30}
\definecolor{green5}{cmyk}{0.25,0,0.75,0.40}
\definecolor{green6}{cmyk}{0.25,0,0.75,0.50}

\definecolor{B02}{cmyk}{0,0.14,0.22,0.12}
\definecolor{B03}{cmyk}{0,0.16,0.26,0.16}
\definecolor{B04}{cmyk}{0,0.19,0.28,0.19}
\definecolor{B05}{cmyk}{0,0.25,0.32,0.25}
\definecolor{B06}{cmyk}{0,0.31,0.36,0.31}
\definecolor{B07}{cmyk}{0,0.37,0.40,0.37}
\definecolor{B08}{cmyk}{0,0.46,0.46,0.46}
\definecolor{B09}{cmyk}{0,0.55,0.52,0.54}
\definecolor{B10}{cmyk}{0,0.69,0.61,0.62}
\definecolor{B11}{cmyk}{0,0.78,0.70,0.68}
\definecolor{B12}{cmyk}{0,0.93,0.85,0.60}
\definecolor{B13}{cmyk}{0.25,1,0.6,0.50}
\definecolor{B14}{cmyk}{0.5,1,0.30,0.40}
\definecolor{B15}{cmyk}{0.75,1,0,0.30}

\definecolor{C02}{cmyk}{0,0.22,0.14,0.12}
\definecolor{C03}{cmyk}{0,0.26,0.16,0.16}
\definecolor{C04}{cmyk}{0,0.28,0.19,0.19}
\definecolor{C05}{cmyk}{0,0.32,0.25,0.25}
\definecolor{C06}{cmyk}{0,0.36,0.31,0.31}
\definecolor{C07}{cmyk}{0,0.40,0.37,0.37}
\definecolor{C08}{cmyk}{0,0.46,0.46,0.46}
\definecolor{C09}{cmyk}{0,0.52,0.55,0.54}
\definecolor{C10}{cmyk}{0,0.61,0.69,0.62}
\definecolor{C11}{cmyk}{0,0.70,0.78,0.68}
\definecolor{C12}{cmyk}{0,0.85,0.93,0.60}
\definecolor{C13}{cmyk}{0.25,0.60,1,0.50}
\definecolor{C14}{cmyk}{0.5,0.30,1,0.40}
\definecolor{C15}{cmyk}{0.75,0,1,0.30}

\begin{document}

\title[Lattice star and acyclic branched polymer vertex exponents in $3d$]{
Lattice star and acyclic branched polymer vertex exponents in $3d$}

\author{S Campbell$^1$ \& EJ Janse van Rensburg$^2\ddagger$}
\address{\sf$^1$Department of Statistics, 
University of Toronto, Toronto, Ontario M3J~4S5, Canada\\}
\address{\sf$^2$Department of Mathematics and Statistics, 
York University, Toronto, Ontario M3J~1P3, Canada\\}
\ead{$\ddagger$\href{mailto:rensburg@yorku.ca}{rensburg@yorku.ca}}
\vspace{10pt}
\begin{indented}
\item[]\today
\end{indented}

\begin{abstract}
Numerical values of lattice star entropic exponents $\gamma_f$, and
star vertex exponents $\sigma_f$, are estimated using parallel implementations 
of the PERM and Wang-Landau algorithms.  Our results show that the numerical 
estimates of the vertex exponents deviate from predictions of the $\eps$-expansion
and confirms and improves on estimates in the literature.

We also estimate the entropic exponents $\gamma_\C{G}$ of a few acyclic 
branched lattice networks with comb and brush connectivities. In particular, 
we confirm within numerical accuracy the scaling relation\cite{D89}
$$ \gamma_{\C{G}}-1 = \sum_{f\geq 1} m_f \, \sigma_f $$
for a comb and two brushes (where $m_f$ is the number of nodes of
degree $f$ in the network) using our independent estimates of $\sigma_f$.
\end{abstract}

%
\vspace{2pc}


%
%

\section{Introduction}
\label{section1}

Lattice self-avoiding walk models of star polymers have been studied since the 
1970s \cite{MM77} and remain of considerable interest in the statistical mechanics 
of polymeric systems.  Numerical simulation of lattice stars stretches back decades
\cite{LWWMG85,WLWG86,BK89,JvRGW89}, and their properties and scaling
exponents have been calculated numerically 
\cite{LWWMG85,WLWG86,WGLW86,BK89,OB91,G97,O02,HNG04,HG11,CJvR21} and 
by using field theoretic approaches \cite{LGZJ77,MF83,MF84,D86,SFLD92,D89}. 
A general survey can be found in references \cite{LSW04,D04}. Recent results 
in reference \cite{DG20} give predictions in models of confined branched and
star polymers.  

\begin{figure}[t]
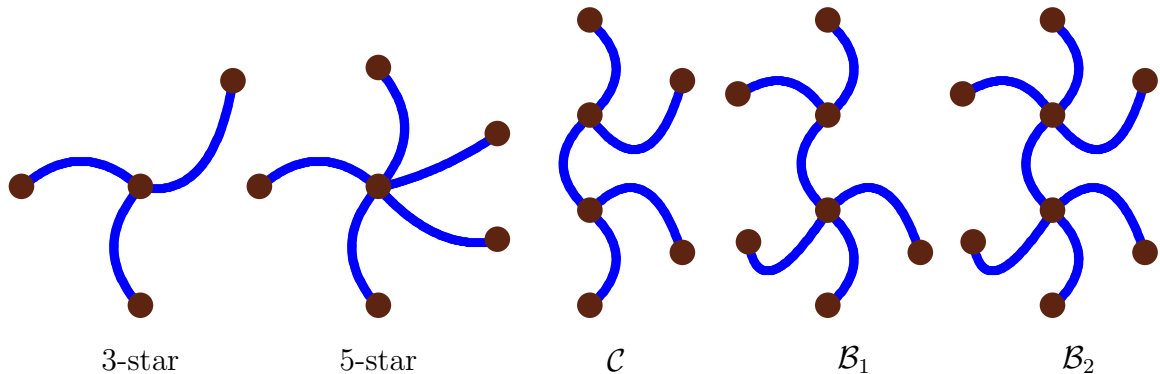

\beginpicture
\setcoordinatesystem units <1.0pt,1.0pt>
\setplotarea x from -140 to 150, y from -20 to 110
\setplotarea x from -100 to 250, y from 0 to 90


\setquadratic

\setplotsymbol ({\scriptsize$\bullet$})
\color{Blue}
\plot -100 0 -110 22.5 -100 45 /
\plot -100 45 -77.5 52.5 -65 85 /
\plot -100 45 -122.5 55 -145 45 /
\color{Sepia}
\multiput {\huge$\bullet$} at -100 0 -100 45 -65 85 -145 45 /
\color{black}
\put {3-star} at -100 -20 

\setcoordinatesystem units <1.0pt,1.0pt> point at 10 0 
\setplotsymbol ({\scriptsize$\bullet$})
\color{Blue}
\plot 0 0 -10 22.5 0 45 /
\plot 0 45 10 67.5 0 90 /
\plot 0 45 22.5 27.5 45 25 /
\plot 0 45 22.5 52.5 45 65 /
\plot 0 45 -22.5 55 -45 45 /
\color{Sepia}
\multiput {\huge$\bullet$} at 0 0 0 45 0 90 45 25 45 65 -45 45 /
\color{Black}
\put {5-star} at 0 -20 

\setcoordinatesystem units <1.0pt,1.0pt> point at 0 0 
\setplotsymbol ({\scriptsize$\bullet$})
\color{Blue}
\plot 70 0 80 18 70 36 60 54 70 72 80 90 70 108 /
\plot 70 36 90 44 105 20 /
\plot 70 72 90 60 105 85 /
\color{Sepia}
\multiput {\huge$\bullet$} at 70 0 70 36 70 72 70 108 105 20 105 85 /
\color{Black}
\put {$\C{C}$} at 80 -20 

\setcoordinatesystem units <1.0pt,1.0pt> point at -90 0 
\setplotsymbol ({\scriptsize$\bullet$})
\color{Blue}
\plot 70 0 80 18 70 36 60 54 70 72 80 90 70 108 /
\plot 70 36 90 44 105 20 /
\plot 70 36 50 14 40 24 /
\plot 70 72 55 85 36 80 /
\color{Sepia}
\multiput {\huge$\bullet$} at 70 0 70 36 70 72 70 108 105 20 40 24 36 80 /
\color{Black}
\put {$\C{B}_1$} at 80 -20 

\setcoordinatesystem units <1.0pt,1.0pt> point at -175 0 
\setplotsymbol ({\scriptsize$\bullet$})
\color{Blue}
\plot 70 0 80 18 70 36 60 54 70 72 80 90 70 108 /
\plot 70 36 90 44 105 20 /
\plot 70 36 50 14 40 24 /
\plot 70 72 90 60 105 85 /
\plot 70 72 55 85 36 80 /
\color{Sepia}
\multiput {\huge$\bullet$} at 70 0 70 36 70 72 70 108 
105 20 40 24 105 85 36 80 /
\color{Black}
\put {$\C{B}_2$} at 80 -20 

\normalcolor
\setlinear
\endpicture
\caption{\textit{Graphs of models examined in this paper.  From the left: 
A $3$-star and a $5$-star, and three branched polymer models, namely 
a comb $\C{C}$, and two brushes $\C{B}_1$ and $\C{B}_2$. 
Vertex exponents are associated with the vertices shown in 
these figures, namely end-vertices of degree $1$, and all other vertices of
degrees bigger than or equal to $3$.}}
\label{models} 
\end{figure}

In this paper we revisit the numerical simulation of $3d$ lattice stars, 
using the Parallel PERM algorithm \cite{G97,PK04,CJvR20}.  Our aims are to update
numerical estimates of the entropic exponent $\gamma_f$ of lattice stars,
and to test the scaling relation \cite{D86}
\begin{equation}
\gamma_{\C{G}}-1 = \sum_{f\geq 1} m_f \, \sigma_f
\label{0}   
\end{equation}
for lattice models of a few uniform branched structures with underlying graph 
or connectivity $\C{G}$ (a comb and two brushes -- see figure \ref{models}), 
where $\gamma_{\C{G}}$ is the entropic exponent of the branched structure, 
and $\sigma_f$ and $m_f$ are the $f$-star vertex exponent and the number 
of nodes of degree $f$ in $\C{G}$, respectively.

An \textit{$f$-star graph} is an acyclic simple graph with one vertex of
degree $f$ (the central vertex), and $f$ vertices of degree $1$ (see figure
\ref{models}).  The $f$-star graph has $f$ \textit{arms} or \textit{branches}
which branch from the central node to their endpoints which are nodes of 
degree $1$.  A \textit{lattice $f$-star} is a lattice embedding of an $f$-star graph 
such that each branch or arm is a self-avoiding walk from the central node 
(of degree $f$) to a node of degree $1$ which is the endpoint of the arm.  
The arms of the embedded lattice star are mutually avoiding but do share 
the central node at their first vertices.  A lattice $f$-star is 
\textit{strictly uniform} if its $f$ arms all have the exact same length.  If the 
longest arm of a lattice star is exactly one step longer than its shortest 
arm, then it is \textit{almost uniform}.  We call lattice $f$-stars of arbitrary 
length $n$ \textit{monodispersed} if they are either strictly, or almost, uniform.  
The \textit{length} of a lattice star is the total number of steps in all 
the arms of the star (that is, the sum of the lengths of the arms).
In the simple cubic (SC) lattice $1\leq f \leq 6$, but it is possible to redefine the
central node such that lattice stars with more than $6$ arms can be embedded.
In the face-centered cubic (FCC) lattice, $1\leq f \leq 12$, and in the body-centered
cubic (BCC) lattice, $1\leq f \leq 8$.

Denote by $c_n$ the number of self-avoiding walks in a lattice 
from the origin, and of length $n$ steps.  The \textit{growth constant} $\mu_d$ 
of self-avoiding walks is given by the limit \cite{HM54,H57,H60}
\begin{equation}
\lim_{n\to\infty} \Sfrac{1}{n} \log c_n = \log \mu_d . 
\label{1}  
\end{equation}
Asymptotically, 
\begin{equation}
c_n = C\, n^{\gamma-1}\, \mu_d^n\, (1+ B\, n^{-\Delta_1} + \cdots)
\label{3A} 
\end{equation}
where $\gamma$ is the \textit{entropic exponent}, and $\Delta_1$ is the
leading confluent correction exponent.  The best estimates of the
entropic exponent in 3d are $\gamma = 1.15698(34)$ 
\cite{SBB11} and $\gamma = 1.15695300(15)$ \cite{C17}.  
The confluence correction exponent was estimated
in reference \cite{CB16} to be $\Delta_1 = 0.528(8)$.

Denote by $s_n^{(f)}$ the number of monodisperse lattice $f$-stars
of length $n$ counted with the central node fixed at the origin.  The growth 
constant of monodispersed lattice $f$-stars is independent of $f$ and is given by
\begin{equation}
\lim_{n\to\infty} \Sfrac{1}{n} \log s_{n}^{(f)} = \log \mu_d ,
\label{1A}  
\end{equation}
where $\mu_d$ is given by equation \Ref{1} \cite{WLWG86,SSW88A,SW89,WS92,SS93}. 
Notice that $s_{nf}^{(f)}$ is the number of strictly uniform $f$-stars 
and $s_{nf+k}^{(f)}$ is the number of almost uniform $f$-stars of 
length $nf{+}k$ for each fixed $k\in\{1,2,\ldots,n-1\}$.  Similar to equation 
\Ref{3A} for $n=mf{+}k$ and $k$ fixed in $\{0,1,2,\ldots, f{-}1\}$
\begin{eqnarray}
s_n^{(f)} \equiv s_{mf+k}^{(f)} 
= C^{(f)}_k\, n^{\gamma_f-1}\,\mu_d^n\,(1+B^{(f)}\,n^{-\Delta_1} + \cdots).
\label{2}  
\end{eqnarray}
The \textit{entropic exponent} $\gamma_f$ is a function of the number 
of arms \cite{D86}.  For values of $k\in\{0,1,\ldots,f{-}1\}$ there are 
persistent parity effects in $s_n^{(f)}$, so that the \textit{amplitudes} 
$C^{(f)}_k$ are functions of $k$ and corresponds to \textit{parity classes}
of monodispersed lattice stars.

\subsection{Lattice star entropic and vertex exponents}

The entropic exponent $\gamma_f$ is related to \textit{vertex exponents}
$\sigma_f$ as shown in equation \Ref{0}.  That is,
\begin{equation}
\gamma_f-1 = \sigma_f + f\,\sigma_1 .
\label{7X}   
\end{equation}
If $f=1$ or $f=2$, then a monodispersed $f$-star is reduced to either
a self-avoiding walk from the origin (when $f=1$), or a self-avoiding
walk with a middle vertex at the origin.  Thus, the number of $1$-stars 
of length $n$ is equal to $c_n$ so that $\gamma_1=\gamma$, and 
by equation \Ref{7X}, using the estimate for $\gamma$ in reference \cite{C17},
\begin{equation}
\sigma_1=(\gamma{-}1)/2=0.0784765(8) .
\label{7Q}  
\end{equation}
The number of strictly uniform $2$-stars of length $2n$ is given 
by $c_{2n}/2$, and the number of almost uniform $2$-stars of 
length $2n{+}1$ is given by $c_{2n+1}$. That is, 
$s_{2n}^{(2)} = c_{2n}/2$ and $s_{2n+1}^{(2)} = c_{2n+1}$.
This shows that $\gamma_2=\gamma$ and by equation \Ref{0},
$\gamma_2-1 = \sigma_2+2\sigma_1$.  This shows that $\sigma_2=0$,
and that $C_0^{(2)}=C/2$ and $C_1^{(2)}=C$.

\begin{table}[t!]
\caption{Estimates of $\gamma_f$ in $3d$}
\begin{indented}
\lineup
\item[]
\begin{tabular}{l|ll|llll}
\br                              
$f$ & This work & Models & Older estimates \cr 
\mr
$3$ & \0\;$1.04282(11)$ & SC, FCC, BCC & $\0\;1.089(1)^1$, $1.0427(7)^4$ \cr
$4$ & \0\;$0.8337(5)$ & SC, FCC, BCC & $\0\;0.879(1)^1$, $0.8355(10)^4$ \cr
$5$ & \0\;$0.5412(8)$ & SC, FCC, BCC & $\0\;0.567(2)^1$, $0.5440(12)^4$ \cr
$6$ & \0\;$0.1726(13)$ & SC, FCC, BCC & $\0\;0.16(1)^1$, $0.1801(20)^4$  \cr

$7$ & $-0.257(3)$ & SC, FCC, BCC & $-0.2520(25)^4$ \cr
$8$ & $-0.757(3)$ & SC, FCC, BCC & $-1.00^2$, $-0.748(3)^4$ \cr
$9$ & $-1.309(4) $ & SC, FCC & $-1.306(5)^4$  \cr
$10$ & $-1.916(5) $ & SC, FCC & $-1.922(7)^4$ \cr
$11$ & $-2.588(6)$ & SC, FCC \cr
$12$ & $-3.305(6)$ & SC, FCC & $-3.35^2$, $-3.4(3)^3$, $-3.296(9)^4$ \cr
\br
\end{tabular}
\begin{tablenotes}
\item 1. References for older estimates: $1$:\cite{BK89}, $2$:\cite{SOKK00}, $3$:\cite{O02}, $4$:\cite{HNG04}
\item 2. SC - simple cubic, FCC - face-centered, BCC - body-centered 
\end{tablenotes}
\end{indented}
\label{tafel1}   
\end{table}

Numerical estimates of $\gamma_f$ can be found in references
\cite{WGLW86,BK89,SOKK00,O02,HNG04} and in this paper we verify
and improve on some of those estimates.  By combining estimates from our
simulations in the SC, FCC and BCC lattices, we obtain the estimates 
in table \ref{tafel1}. .

\begin{table}[h!]
\caption{Vertex exponents $\sigma_f$ of $f$-stars in $3d$}
\begin{indented}
\lineup
\item[]\begin{tabular}{@{}*{6}{l}}
\br                              
$f$ & \;\cite{WGLW86} & \;\cite{BK89} & \;\cite{SOKK00} & \;\cite{HNG04} & This work \cr 
\mr
$\sigma_1$ & $\0\;-$             & $\0\;0.0855(5)$ & $\0\;-$ & $\0\,0.07865(10)$ & $\;-$\0  \cr
$\sigma_3$ & $-0.19(3)$  & $-0.1675(5)$     & $-0.216$ & $-0.1927(7)$ &  $-0.19313(11)$  \cr
$\sigma_4$ & $-0.44(3)$  & $-0.463(1)\0$    & $\0\;-$     & $-0.4784(10)$ &  $-0.4802(5)$  \cr
$\sigma_5$ & $-0.85(5)$  & $-0.8605(5)$     & $\0\;-$     & $-0.8484(12)$ &  $-0.8512(8)$  \cr
$\sigma_6$ & $-1.28(5)$  & $-1.353(7)\0$    & $-1.401$ & $-1.2908(20)$ &  $-1.2983(13)$  \cr
\br
\end{tabular}
\end{indented}
\label{table00}  
\end{table}

Accurate estimates of $\sigma_f$ in 3d can be obtained from the data 
in table \ref{tafel1}, equation \Ref{7X}, and the estimate of $\sigma_1$ 
in equation \Ref{7Q}.  The results are shown for $3\leq f \leq 6$ in 
the last column of table \ref{table00}, where earlier results (obtained
from estimates of $\gamma_f$ in those references and equations \Ref{7X}
and \Ref{7Q}).

Studies using renormalization group methods and the $\eps$-expansion 
\cite{D86,SFLD92} gives
\begin{equation}
\gamma_f = \sigma_f + f \,\sigma_1 = \nu ( \eta_f - f \,\eta_2 ) .
\end{equation}
The numbers $\eta_f$ were calculated to $O(\eps^3)$ \cite{SFLD92}.  To second 
order in $\eps$,
\begin{equation}
\eta_f = - \Sfrac{\eps}{8}\,f\,(f-1)\, \LB
1 - \Sfrac{\eps}{32} (8\,f - 25) + O(\eps^2) \RB .
\label{7A}  
\end{equation}
These $\eps^2$ estimates are best for small values of $f\leq 4$, and has low
fidelity for $f\geq 5$.  A second order $\eps$-expansion for $\gamma$ \cite{LGZJ77}
(and thus for $\sigma_1$) gives the following second order $\eps$-expansion for 
$\sigma_f$ \cite{D86,D89}:
\begin{equation}
\sigma_f = \Sfrac{\epsilon}{16}\,f\,(2-f)
 + \Sfrac{\epsilon^2}{512}\,f\,(f-2)\,(8\,f-21)
  + O(\epsilon^3).
\label{8}  
\end{equation}
In reference \cite{SFHFB03} equation \Ref{7A} was extended by calculating terms up to
order $\eps^4$.  We list predictions using the $\eps$-expansion for $\sigma_f$ 
in three dimensions to order $\eps^k$ for $k=1,2,3,4$ in table \ref{table000}: 
The order $\eps^1$ and $\eps^2$ approximations are obtained from 
equation \Ref{8}, while the order $\eps^3$ and $\eps^4$ estimates follow
from reference \cite{SFHFB03}.  The $[3/2]$ Pad\'e approximation was 
determined by a Borel resummation of the order $\eps^4$ expansion 
and then using the Pad\'e approximant to recalculate the exponents.  The results 
are shown in the sixth column. The Pad\'e and $\eps^1$ results approximate
the $\sigma_f$ reasonably for $f\leq 4$, but even these deviate from the 
numerical data for $f=5$ and $f=6$.

\begin{table}[h!]
\caption{Vertex exponents: $\eps$-expansions in 3 dimensions}
\begin{indented}
\lineup
\item[]\begin{tabular}{@{}*{8}{l}}
\br                              
$f$ & $\epsilon$ \cite{LGZJ77} & $\epsilon^2$ \cite{LGZJ77}
 & $\epsilon^3$ \cite{SFHFB03} & $\epsilon^4$ \cite{SFHFB03}
 & Pad\'e[3/2]  & This work \cr 
\mr
$\sigma_1$ &  $\0\;0.0625$ & $\0\;0.0879$  
                   & $\0\;0.0556$  & $\0\00.1425$ & $\0\;0.0798$ 
                   & $-$ \cr
$\sigma_3$ &  $-0.1875$     & $-0.1699$      
                   & $-0.2265$ & $\;-0.1094$         & $-0.2026$
                   &  $-0.19313(11)$ \cr
$\sigma_4$ &  $-0.5\0\0\0$  & $-0.3281$      
                   & $-0.9037$ & $\0\00.9925$      & $-0.5079$
                   &  $-0.4802(5)$  \cr
$\sigma_5$ &  $-0.9375$     & $-0.3809$      
                   & $-2.5321$ & $\0\06.0673$      & $-0.8964$
                   & $-0.8512(8)$  \cr
$\sigma_6$ &  $-1.5\0\0\0$  & $-0.2344$      
                   & $-5.8322$ & $\020.1822$       & $-1.3594$
                   &  $-1.2983(13)$    \cr
\br
\end{tabular}
\end{indented}
\label{table000}   
\end{table}

Asymptotically $\gamma_f$ and $\sigma_f$ scales with $f$ as \cite{WP86,O02}
\begin{equation}
\gamma_f - 1 \sim \sigma_f + f\,\sigma_1 \sim - f^{3/2} .
\label{eqn12}  
\end{equation}
Comparison to equation \Ref{8} suggests that the $\epsilon$-expansion
should break down quickly with increasing $f$, as the order $\eps^n$ term
is seen to grow as $O(f^{n+1})$.  Moreover, it cannot be improved by 
calculating ever higher order corrections, as the coefficients increase
quickly in magnitude with increasing $f$.  Resummation techniques do give 
improved estimates, but are limited for large $f$, and by the increasing 
complexity of calculating higher order terms in longer $\eps$-expansions.  
See, for example, chapter 16 in reference \cite{KS-F01}.

\begin{figure}[t]
\centering
\begin{minipage}{0.33\textwidth}
   \centering
\scalebox{0.5}{
    \begin{tikzpicture}
\color{Blue}
    \draw[step=1cm,Tan,very thin] (-4,-4) grid (5,4);
    \draw [line width=1.5mm](-2,-2) -- (-2,0) -- (-1,0) ;
    \draw [line width=1.5mm](-2,2) -- (-1,2) -- (-1,0) ;
    \draw [line width=1.5mm](-1,0) -- (2,0);
    \draw [line width=1.5mm] (2,0) -- (2,-2) -- (3,-2);
    \draw [line width=1.5mm] (2,0) -- (3,0) -- (3,1) -- (4,1);
    \node at (-1,0) [circle,fill=blue,scale=1] {};
    \node at (-2,-2) [circle,fill=blue,scale=1] {};
    \node at (-2,2) [circle,fill=blue,scale=1] {};
    \node at (2,0) [circle,fill=blue,scale=1] {};
    \node at (3,-2) [circle,fill=blue,scale=1] {};
    \node at (4,1) [circle,fill=blue,scale=1] {};
    \node at (-2,-1) [circle,fill=black,scale=0.75]{};
    \node at (-2,0) [circle,fill=black,scale=0.75]{};
    \node at (-1,1) [circle,fill=black,scale=0.75]{};
    \node at (-1,2) [circle,fill=black,scale=0.75]{};
    \node at (0,0) [circle,fill=black,scale=0.75]{};
    \node at (1,0) [circle,fill=black,scale=0.75]{};
    \node at (3,0) [circle,fill=black,scale=0.75]{};
    \node at (3,1) [circle,fill=black,scale=0.75]{};
    \node at (2,-1) [circle,fill=black,scale=0.75]{};
    \node at (2,-2) [circle,fill=black,scale=0.75]{};
    \end{tikzpicture}
    }
    \caption{$\mathcal{C}$}
    \label{Comb1}
\end{minipage} \hfill
\begin{minipage}{0.33\textwidth}
 \centering
\scalebox{0.5}{
    \begin{tikzpicture}
\color{Blue}
    \draw[step=1cm,Tan,very thin] (-4,-4) grid (5,4);
    \draw [line width=1.5mm](-2,-2) -- (-2,0) -- (-1,0) ;
    \draw [line width=1.5mm](-2,2) -- (-1,2) -- (-1,0) ;
    \draw [line width=1.5mm](-1,0) -- (2,0);
    \draw [line width=1.5mm] (2,0) -- (2,-2) -- (3,-2);
    \draw [line width=1.5mm] (2,0) -- (4,0) -- (4,1);
    \draw [line width=1.5mm] (2,0) --  (2,3);
    \node at (-1,0) [circle,fill=blue,scale=1] {};
    \node at (-2,-2) [circle,fill=blue,scale=1] {};
    \node at (-2,2) [circle,fill=blue,scale=1] {};
    \node at (2,0) [circle,fill=blue,scale=1] {};
    \node at (3,-2) [circle,fill=blue,scale=1] {};
    \node at (4,1) [circle,fill=blue,scale=1] {};
    \node at (2,3) [circle,fill=blue,scale=1] {};
    \node at (-2,-1) [circle,fill=black,scale=0.75]{};
    \node at (-2,0) [circle,fill=black,scale=0.75]{};
    \node at (-1,1) [circle,fill=black,scale=0.75]{};
    \node at (-1,2) [circle,fill=black,scale=0.75]{};
    \node at (0,0) [circle,fill=black,scale=0.75]{};
    \node at (1,0) [circle,fill=black,scale=0.75]{};
    \node at (2,1) [circle,fill=black,scale=0.75]{};
    \node at (2,2) [circle,fill=black,scale=0.75]{};
    \node at (2,-1) [circle,fill=black,scale=0.75]{};
    \node at (2,-2) [circle,fill=black,scale=0.75]{};
    \node at (3,0) [circle,fill=black,scale=0.75]{};
    \node at (4,0) [circle,fill=black,scale=0.75]{};
    \end{tikzpicture}
    }
    \caption{$\mathcal{B}_1$}
    \label{Brush1}
\end{minipage}\hfill
\begin{minipage}{0.33\textwidth}
   \centering
\scalebox{0.5}{
    \begin{tikzpicture}
\color{Blue}
    \draw[step=1cm,Tan,very thin] (-4,-4) grid (5,4);
    \draw [line width=1.5mm](-1,-3) -- (-1,0);
    \draw [line width=1.5mm](-2,2) -- (-1,2) -- (-1,0) ;
    \draw [line width=1.5mm](-1,0) -- (2,0);
    \draw [line width=1.5mm] (2,0) -- (2,-2) -- (3,-2);
    \draw [line width=1.5mm] (-1,0) -- (-2,0) -- (-2,-1) -- (-3,-1);
    \draw [line width=1.5mm] (2,0) -- (4,0) -- (4,1);
    \draw [line width=1.5mm] (2,0) --  (2,3);
    \node at (-1,0) [circle,fill=blue,scale=1] {};
    \node at (-1,-3) [circle,fill=blue,scale=1] {};
    \node at (-2,2) [circle,fill=blue,scale=1] {};
    \node at (2,0) [circle,fill=blue,scale=1] {};
    \node at (3,-2) [circle,fill=blue,scale=1] {};
    \node at (4,1) [circle,fill=blue,scale=1] {};
    \node at (2,3) [circle,fill=blue,scale=1] {};
    \node at (-3,-1) [circle,fill=blue,scale=1] {};
    \node at (-2,-1) [circle,fill=black,scale=0.75]{};
    \node at (-2,0) [circle,fill=black,scale=0.75]{};
    \node at (-1,-2) [circle,fill=black,scale=0.75]{};
    \node at (-1,-1) [circle,fill=black,scale=0.75]{};
    \node at (-1,1) [circle,fill=black,scale=0.75]{};
    \node at (-1,2) [circle,fill=black,scale=0.75]{};
    \node at (0,0)  [circle,fill=black,scale=0.75]{};
    \node at (1,0) [circle,fill=black,scale=0.75]{};
    \node at (3,0) [circle,fill=black,scale=0.75]{};
    \node at (4,0) [circle,fill=black,scale=0.75]{};
    \node at (2,-2) [circle,fill=black,scale=0.75]{};
    \node at (2,-1) [circle,fill=black,scale=0.75]{};
    \node at (2,1) [circle,fill=black,scale=0.75]{};
    \node at (2,2) [circle,fill=black,scale=0.75]{};
    \end{tikzpicture}
    }
    \caption{$\mathcal{B}_2$}
    \label{Brush2}
\end{minipage}\hfill
\end{figure} 

\subsection{Uniform acyclic branched structures (uniform trees)}
\label{section11}  

Models of branched polymeric structures are \textit{lattice networks} with 
connectivities or topologies denoted by $\C{G}$.  A lattice network consists of
\textit{branches} which are self-avoiding walks joining \textit{vertices} 
of degrees equal to $1$ or bigger than or equal to $3$.  The underlying 
connectivity of a lattice network is denoted by a graph.  This may be, for 
example, one of the cases shown in figure \ref{models}. Lattice $f$-stars 
are examples of lattice networks of fixed connectivity, and so 
are the other cases shown in figure \ref{models}.  These cases include a 
\textit{comb} $\C{C}$ (figure \ref{Comb1}), and two \textit{brushes},
$\C{B}_1$ (figure \ref{Brush1}) and $\C{B}_2$ (figure \ref{Brush2}).

A lattice network is \textit{strictly uniform} if all the branches are 
self-avoiding walks of the same length.  If a strictly uniform lattice network 
of connectivity $\C{G}$ has $b$ branches, then the total number of such 
networks, equivalent under translations in the lattice,  is denoted by 
$c_{n}(\C{G})$, and it is generally accepted that
\begin{equation}
c_{n}(\C{G}) = C_\C{G}\, n^{\gamma_\C{G} - 1}\, \mu_d^{n}\,(1+o(1)) .
\label{14CX}  
\end{equation}
The growth constant $\mu_d$ is equal to that of self-avoiding walks 
\cite{SS93,SW89,SSW88A,WS92}.   The relation of the entropic exponent
$\gamma_\C{G}$ for a general network connecticity $\C{G}$ and star
vertex exponents $\sigma_f$ is given by the relation
\begin{equation}
\gamma_\C{G} - 1 = \sum_{f\geq 1} m_f \, \sigma_f - c(\C{G}) \, d\nu,
\label{14C}  
\end{equation}
where $m_f$ is the number of vertices of degree $f$, and where $c(\C{G})$ is
the cyclomatic index (the number of independent cycles) in the network
\cite{D86,D89} (see equation \Ref{0} for the case when $c(\C{G})=0$).
The models in figure \ref{models} are acyclic, and by the above
\begin{eqnarray}
\gamma_{\C{C}} - 1 & = &  4\,\sigma_1 + 2\, \sigma_3 , \nonumber \\ 
\gamma_{\C{B}_1} - 1 & = & 5\,\sigma_1 + \sigma_3 + \sigma_4,
\label{15CC}  
 \\
\gamma_{\C{B}_2} - 1 & = & 6\,\sigma_1 + 2\, \sigma_4 . \nonumber 
\end{eqnarray}
The underlying assumption is that the vertex exponents $\sigma_f$ are 
independent of the connectivity of the uniform branched polymer so that 
the presence of other nodes of given degree does not affect its value.
For large $n$ this is a reasonable assumption since the distance between 
any two branch points in the lattice graph increases as $O(n^{\nu_d})$ 
(where $\nu_d$ is the self-avoiding walk metric exponent in dimension $d$).

\begin{table}[t]
\caption{$\gamma_{\mathcal{G}}{-}1$ for lattice networks in the cubic lattice}
\begin{indented}
\lineup
\item[]
\scalebox{1.0}{
\begin{tabular}{c | @{}*{4}{c}}
\br
$\C{G}$ & $\0\0\eps^1$ & Pad\'e[3/2] & Eqn \Ref{15CC} & This work \cr 
\mr
$\C{C}$     & $\0-0.1250$ & $-0.0860$ & $-0.07127(20)$ & $-0.0731(18)$ \cr
$\C{B}_1$ & $\0-0.3750$ & $-0.3115$ & $-0.28000(32)$ &  $-0.2896(69)$ \cr
$\C{B}_2$ & $\0-0.6250$ & $-0.5370$ & $-0.48874(44)$ &  $-0.5065(58)$ \cr
\br
\end{tabular}
}
\end{indented}
\label{table0000}   
\end{table}

Our results for lattice networks are shown in table \ref{table0000}.  In three
dimensions the predicted $\eps$-expansion results (at order $\eps^1$)
deviate from the numerical results.  Higher order $\eps$-expansion estimates
do not improve these.  The prediction using the Pad\'e[3/2] estimates improves
on the estimates $\eps^1$.  The predictions by equation \Ref{15CC} are 
obtained by using our best estimates for $\sigma_f$ in table \ref{table000}, 
while the estimates in the final column were obtained by analysing our data 
for networks in section \ref{S51}.

\def\stepb#1#2#3#4{\color{Blue}
                                   \plot #1 #2 #3 #4  /
                                   \color{black} 
                                   \multiput {\large$\bullet$} at #1 #2 #3 #4 /
                                   }
\def\stepr#1#2#3#4{\color{Red}
                                   \plot #1 #2 #3 #4  /
                                   \color{black} 
                                   \multiput {\large$\bullet$} at #1 #2 #3 #4 /
                                   }
\def\stepg#1#2#3#4{\color{DarkGreen}
                                   \plot #1 #2 #3 #4  /
                                   \color{black} 
                                   \multiput {\large$\bullet$} at #1 #2 #3 #4 /
                                   }
\def\putnum#1#2#3{\put {\scalebox{0.65}{#1}} at #2 #3 }

\begin{figure}[t]
\beginpicture
\setcoordinatesystem units <2pt,2pt> point at 0 0 
\setplotarea x from -110 to 50, y from -60 to 60
\setplotarea x from -50 to 50, y from -50 to 60

\color{Tan}
\grid 10 11 

\setplotsymbol ({\scriptsize$\bullet$})

\stepb{0}{0}{10}{0}         \putnum{1}{5}{-3}
\stepb{10}{0}{10}{-10}    \putnum{4}{13}{-5}
\stepb{10}{-10}{20}{-10} \putnum{7}{15}{-13}
\stepb{20}{-10}{20}{0}    \putnum{10}{23}{-5}
\stepb{20}{0}{30}{0}       \putnum{13}{25}{3}
\stepb{30}{0}{30}{10}       \putnum{16}{33}{5}
\stepb{30}{10}{40}{10}    \putnum{19}{35}{13}
\stepb{40}{10}{40}{20}   \putnum{22}{43}{15}
\stepb{40}{20}{30}{20}   \putnum{25}{35}{23}
\stepb{30}{20}{30}{30}   \putnum{28}{27}{25}

\stepr{0}{0}{-10}{0}    \putnum{2}{-5}{-3}
\stepr{-10}{0}{-20}{0}  \putnum{5}{-15}{-3}
\stepr{-20}{0}{-20}{-10}  \putnum{8}{-23}{-5}
\stepr{-20}{-10}{-30}{-10}  \putnum{11}{-25}{-13}
\stepr{-30}{-10}{-30}{-20}  \putnum{14}{-33}{-15}
\stepr{-30}{-20}{-20}{-20}  \putnum{17}{-25}{-23}
\stepr{-20}{-20}{-10}{-20}  \putnum{20}{-15}{-23}
\stepr{-10}{-20}{-10}{-30}  \putnum{23}{-7}{-25}
\stepr{-10}{-30}{0}{-30}    \putnum{26}{-5}{-33}

\stepg{0}{0}{0}{10}   \putnum{3}{-3}{5}
\stepg{0}{10}{-10}{10}  \putnum{6}{-5}{13}
\stepg{-10}{10}{-10}{20}  \putnum{9}{-13}{15}
\stepg{-10}{20}{-20}{20}  \putnum{12}{-15}{23}
\stepg{-20}{20}{-20}{30}  \putnum{15}{-23}{25}
\stepg{-20}{30}{-10}{30}  \putnum{18}{-15}{33}
\stepg{-10}{30}{0}{30}   \putnum{21}{-5}{33}
\stepg{0}{30}{0}{40}   \putnum{24}{3}{35}
\stepg{0}{40}{10}{40}   \putnum{27}{5}{43}

\color{black} 
\put {$\bullet$} at 0 0 
\setplotsymbol ({.})
\circulararc 360 degrees from 3 0 center at 0 0

\setdots <3pt>
\setplotsymbol ({\scalebox{0.5}{$\bullet$}})
\plot 0 -30 0 -18 /  \plot 0 -30 12 -30 /   \plot 0 -30 0 -42 /
\plot 30 30 42 30 /  \plot 30 30 30 42 /  \plot 30 30 18 30 /
\plot 10 40 22 40 /  \plot 10 40 10 52 / \plot 10 40 10 28 /

\setsolid
\setplotsymbol ({.})
\put {\scalebox{1.1}{1}} at 22 15
\circulararc 360 degrees from 22 19 center at 22 15
\put {\scalebox{1.1}{2}} at -20 -30
\circulararc 360 degrees from -20 -34 center at -20 -30
\put {\scalebox{1.1}{3}} at -2 20
\circulararc 360 degrees from -2 24 center at -2 20

\color{black}
\normalcolor
\endpicture
\caption{\textit{Growing a lattice $3$-star by PERM.  The three branches are 
labeled by $\{1,2,3\}$ and the growth point cycles with each step from one branch
to the next.  The first step is added to the origin to start growing
the first arm, then the second step and third steps are added to initiate
the second and third arms.  The growth point is then situated at the endpoint
of the first branch and the fourth step is added here.  The growth point moves to
the end point of the second branch, and the fifth step is added here, and so on.
Eventually, step $m$ is appended to arm $k=m\,\hbox{mod}\,f$.
This elementary move is implemented using Rosenbluth dynamics \cite{RR55}, 
which with enrichment and pruning \cite{G97} gives the PERM algorithm
for lattice stars.  In some of our simulations this algorithm was implemented 
using the parallel implementation in reference \cite{CJvR20}.  Notice that only 
monodisperse (uniform or almost uniform) lattice stars are sampled.}}
\label{figureZ}  
\end{figure}

\section{Parallel PERM sampling of lattice stars}
\label{S2}  

Our Parallel PERM simulations of self-avoiding $f$-stars in the SC, FCC and 
BCC lattices were done by initialising $f$-stars with their central nodes at 
the origin and then growing branches by appending steps at the endpoints 
of arms one at a time while cycling through the $f$ arms of the star.  
For example, in figure \ref{figureZ} the steps in a $3$-star are labelled in 
the order they were appended.  The steps along the arm labeled $1$, 
were added first, fourth, seventh, tenth, and so on, giving the sequence 
of labels $(1,4,7,10,13,16,19,22,25,28)$ along this arm. The second arm 
is grown starting at step $2$, and so on.  Appending a step with label 
$fm$ gives a uniform $f$-star, while the other cases give almost 
uniform stars.  That is, the algorithm samples in the state space of monodisperse $f$-stars producing approximate counts of the number 
of monodispersed $f$-stars.  Using this implementation, $f$-stars were 
sampled in the SC to $f=6$, the FCC to $f=12$ and the BCC to $f=8$.

There are $f!$ ways in which to grow a uniform $f$-star, and $k!(f{-}k)!$ 
ways to grow an almost uniform $f$-star of length $f m{+}k$.  That is, 
PERM estimates the quantity $u_n^{(f)} = k!(f{-}k)!\,s_n^{(f)}$ 
if $n=fm{+}k$.  By equation \Ref{2},
\begin{equation}
u_n^{(f)} = U^{(f)}\, n^{\sigma_f + f\,\sigma_1}\,\mu_d^n\,(1+o(1)),
\label{14}  
\end{equation}
and $U^{(f)}$ is related to $C^{(f)}_k$ in equation \Ref{2}  by
\begin{equation}
U^{(f)} = k!(f{-}k)!\, C^{(f)}_k .
\label{15} 
\end{equation}
The \textit{amplitudes} of the parity classes of monodisperse $f$-stars,
$C^{(f)}_k$, can be estimated for each $k$ by first estimating $U^{(f)}$ 
from our data. 

We also used Parallel PERM to sample $f$-stars with $f>6$ in the SC lattice.  
These are grown by initiating each arm at a vertex with coordinates 
$(\pm a, \pm b, \pm c)$ where the signs are chosen and the coordinates 
permutated randomly.  We used the values $(a,b,c)=(0,1,2)$ to sample 
$f$-stars with $7\leq f\leq 12$ (so that each arm is initiated on a sphere 
of radius $\sqrt{5}$ centered at the origin in the SC lattice).  In our
simulations we used a 32 bit implementation of the Mersenne 
twistor \cite{MN98} to generate random numbers.

\subsection{Results from lattice stars in the SC lattice}

\begin{figure}[t]
\centering
\includegraphics[width=0.75\textwidth]{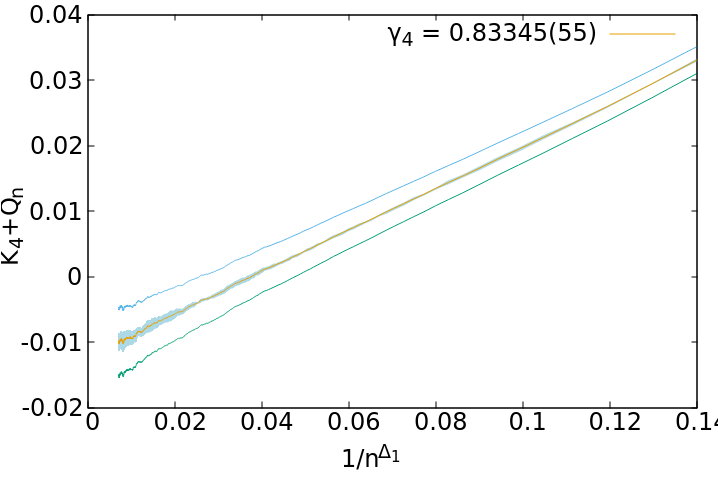}
\caption{\textit{$K_4+Q_n$ for $4$-stars in the cubic lattice 
plotted for $\gamma_4=0.83343(55)$.  $K_4$ was chosen such
that the middle curve passes through zero when $n=500$. The middle curve
corresponds to the raw data with width equal to error bars on the data.  The 
curves on either side are plots of $K_4+Q_n$, but now with $\gamma_4$ taking
its values at the limits of its error bar. The top curve is slightly convex, and
the bottom curve is slightly concave.}}
\label{figure6}
\end{figure}

For cubic lattice $f$-stars equation \Ref{15} becomes
\begin{equation}
u_n^{(f)} = U^{(f)}\,n^{\gamma_f-1}\,\mu_d^n
(1+B^{(f)}n^{-\Delta_1}+ \cdots) .
\end{equation}
Dividing both sides by $n^{\gamma_f-1}\,\mu_d^n$ and taking logarithms gives 
\begin{equation}
\hspace{-2.5cm}
Q_n = \log \LB \frac{u_n^{(f)}}{\mu_3^n\, n^{\gamma_f-1}} \RB 
\simeq S^{(f)} + \log (1 + B^{(f)}\,n^{-\Delta_1}) + \cdots 
= S^{(f)} + B_0^{(f)}\, n^{-\Delta_1} + \cdots.
\label{30x}
\end{equation}
Estimates of $\gamma_f$ are obtained using the best estimates 
$\mu_3=4.684039931(27)$ \cite{C13} of the cubic lattice self-avoiding walk 
growth constant and $\Delta_1 = 0.528(8)$ \cite{CB16} in equation \Ref{30x}.
If the correct value of $\gamma_f$ is inserted on the left hand side, then
$Q_n$ is to leading order a linear function of $n^{-\Delta_1}$.  Thus, 
we determine the best estimate for $\gamma_f$, assuming the model
\begin{equation}
Q_n = A_f + B_f\, n^{-\Delta_1} + C\, n^{-1}
\label{19X}
\end{equation}
where an analytic correction is included.  Plotting $Q_n$ against 
$n^{-\Delta_1}$ can be interpolated on the value of $\gamma_f$ to obtain 
that best estimate where the graph is a straight line, except perhaps
at the smallest values of $n$.   In figure \ref{figure6} this is shown 
for cubic lattice $4$-stars by plotting $K_4+Q_n$ against $n^{-\Delta_1}$
and with $K_4$ chosen so that the curve passes through zero at $n=500$.  
This simulation included $1.1334\times 10^{9}$ realised parallel PERM 
tours along $4$ parallel sequences of $4$-stars to total length $n=12,\!000$,
and the graph straightens out when $\gamma_f=0.83345(55)$.
The confidence interval is determined by find those values of $\gamma_f$
where the top curve is clearly convex, and the bottom curve is clearly
concave.

\begin{figure}[t]
\centering
\includegraphics[width=0.75\textwidth]{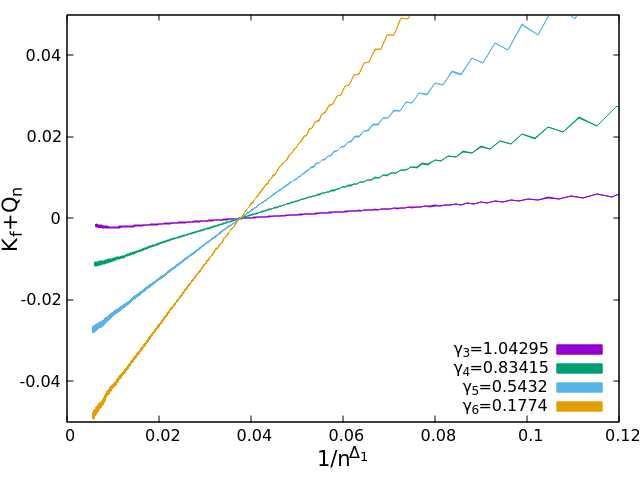}
\caption{\textit{$K_f{+}Q_n$ plotted against $1/n^{\Delta_1}$ with $K_f$ chosen
such that the curves pass through zero when $n=500$.  In this curves straightened
and gives the values for $\gamma_f$ shown in the legend of the plot.}}
\label{figure7}
\end{figure}

For $f$-stars in the SC lattice with $3\leq f\leq 6$ the analytic correction 
in equation \Ref{19X} proved to be negligible, even for small values of $n$, 
and so it could be ignored.  In figure \ref{figure7} our data are plotted for 
these $f$-stars using our best estimates of $\gamma_f$.  Our results are 
collected in table \ref{tableE}, together with the parameters of our simulations.

\begin{table}[h!]
\caption{Estimates of $\gamma_f$ in the SC lattice}
\begin{indented}
\lineup
\item[]\begin{tabular}{@{}*{5}{c}}
\br                              
$f$& Length & Tours & $\gamma_f$   \cr 
\mr
$3$     & $12,\!000$ & $1.1303\times 10^9$       & $\0\;1.0423(3)    $   \cr 
$4$     & $12,\!000$ & $1.1334\times 10^9$       & $\0\;0.8335(6)    $   \cr 
$5$     & $12,\!000$ & $1.1505\times 10^9$       & $\0\;0.5411(9)    $   \cr 
$6$     & $12,\!000$ & $8.2231\times 10^8$       & $\0\;0.1738(18)    $   \cr 
$7$     & $12,\!005$ & $6.2800\times 10^8$       & $-0.253(4)     $   \cr 
$8$     & $12,\!000$ & $6.0800\times 10^8$       & $-0.754(4)     $   \cr 
$9$     & $12,\!006$ & $5.5200\times 10^8$       & $-1.306(5)     $   \cr 
$10$   & $12,\!000$ & $6.0300\times 10^8$       & $-1.912(6)     $   \cr 
$11$   & $13,\!200$ & $6.3100\times 10^8$       & $-2.581(8)     $   \cr 
$12$   & $14,\!400$ & $6.3300\times 10^8$       & $-3.301(9)     $   \cr 
\br
\end{tabular}
\end{indented}
\label{tableE}  
\end{table}

\begin{figure}[t]
\centering
\includegraphics[width=0.45\textwidth]{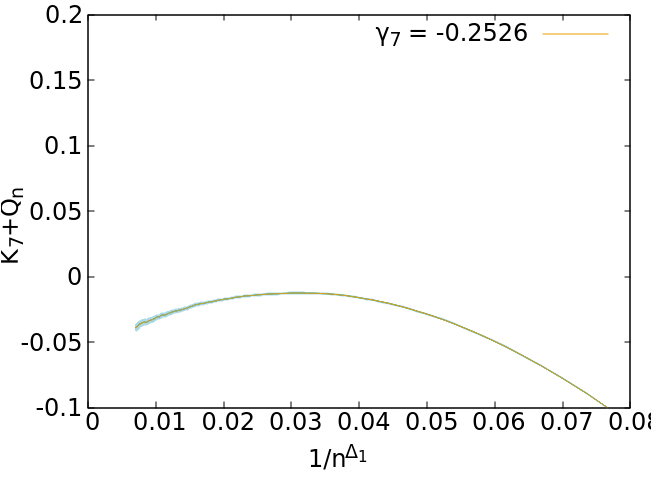}
\includegraphics[width=0.45\textwidth]{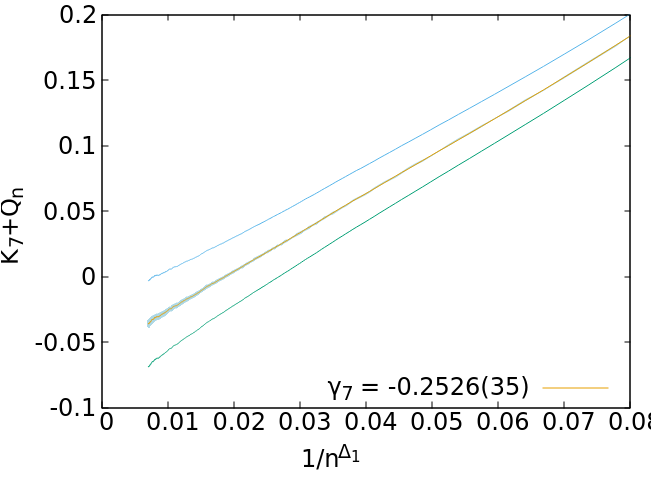}
\caption{\textit{$K_7{+}Q_n$ plotted against $1/n^{\Delta_1}$
with $\gamma_7=-0.2526(35)$ using cubic lattice data for $7$-stars with
arms initiated on lattice sites on a sphere of radius $\sqrt{5}$ centered at the origin.  
Left panel:  Plot without taking account of the analytic correction in the data.  
Right panel:  The same data but now with an analytic correction included and
with $K_7$ chosen such that the middle curve passes through zero when $n=500$.  
In the right panel the top and bottom curves correspond to the value of $\gamma_7$ 
at the limits of its error bars.  The plots are for data with $n\geq 120$, and 
the estimate of $\gamma_7$ were extracted from data with $n\geq 200$.}}
\label{figure8}
\end{figure}

A different approach was followed to extract $\gamma_f$ for $6<f\leq 12$.   
First note that arms of the stars were seeded at lattice points a distance 
$\sqrt{5}$ from the origin, increasing the length of each arm by this amount, 
resulting in a stronger correction term of order $n^{-2\Delta_1}$ in 
equation \Ref{30x}.  Since $2\Delta_1 \approx 1$ the numerical effect is
evident in a very strong analytic correction in our data. This effect was 
already noted in reference \cite{HNG04}, and to compensate for 
this a modified approach was used to estimate of $\gamma_f$ from 
cubic lattice data with $f>6$.  Thus, we proceeded by writing 
equation \Ref{30x} in the form
\begin{equation}
P_n =  \log \LB \frac{u_n^{(f)}}{\mu_3^n} \RB
= (\gamma_f{-}1)\,\log n + a_f + b_f \, n^{-\Delta_1}+c_f\, n^{-1} .
\end{equation}
Since parity effects in our data are of period $2f$, substract $P_{n-2f}$ from $P_n$ 
and then expand the result in $n$ to obtain the model
\begin{equation}
R_n = P_n - P_{n-2f}
= 2f\,(\gamma_f{-}1)\,(n^{-1}+ f\,n^{-2}) + b_f^\prime \, n^{-1-\Delta_1} 
 + c_f^\prime\,n^{-2} .
\end{equation}
Linear least squares analysis using this model for $n>n_{min}=200$ gave
good estimates of $\gamma_f$.  For example, if $f=7$ this gives
$\gamma_7=-0.2526\ldots$ (this compares well with the estimate obtained 
from the Domb-Joyce model in reference \cite{HNG04}).  In addition,
we found that $c_7^\prime=35.57\ldots$.  The effect of the analytic 
correction can be checked by plotting $K_7{+}Q_n$ against
$1/n^{\Delta_1}$ without the analytic correction.  This is shown in 
the left panel of figure \ref{figure8}.  The result is a strongly curved graph, 
which straightens when the correction is included (right panel).  
In the right panel we also determine an error bar on the estimate 
of $\gamma_7$ using the same approach as before.  This gives
$\gamma_7=-0.2526(35)$.  Estimates for $7\leq f \leq 12$ are shown
in table \ref{tableE}.

\subsection{Results: FCC and BCC lattice stars}

Simulations in the FCC and BCC lattices were performed similarly to those in the cubic
lattice.  In the FCC we are able to grow stars with central node at the origin
for $3\leq f\leq 12$, and in the BCC lattice for $3\leq f\leq 8$.  In each of these
lattices good estimates of the growth constant $\mu_3$ were needed (see equation
\Ref{30x}).  The most precise estimates (extrapolated from exact 
enumeration data) were obtained in reference \cite{SBB11} and are
\begin{equation}
\mu_3 =
\cases{
10.037075(20), & (FCC); \cr
6.530520(20), & (BCC).
}
\end{equation}
Using these estimates in equation \ref{30x} proved that they are accurate 
enough to extract accurate estimates of the $f$-star exponents $\gamma_f$. 
Our analysis proceeded as in figures \ref{figure6} and \ref{figure7}.  
Details of our results and estimates of $\gamma_f$ are shown in 
tables \ref{tableEE} and \ref{tableEF}.

\begin{table}[h!]
\caption{Estimates of $\gamma_f$ in the FCC lattice}
\begin{indented}
\lineup
\item[]\begin{tabular}{@{}*{5}{c}}
\br                              
$f$& Length & Tours & $\gamma_f$ (FCC)  \cr 
\mr
$3$     & $12,\!000$ & $2.026 \times 10^8$ & $1.04290(12)$   \cr 
$4$     & $12,\!000$ & $2.800 \times 10^8$ & $0.8343(15)$     \cr 
$5$     & $12,\!000$ & $2.150 \times 10^8$ & $0.5415(25)$  \cr 
$6$     & $12,\!000$ & $2.102 \times 10^8$ & $0.1712(28)$  \cr 
$7$     & $14,\!000$ & $2.866 \times 10^8$ & $-0.258(4)$   \cr 
$8$     & $14,\!000$ & $2.836 \times 10^8$ & $-0.757(5)$   \cr 
$9$     & $14,\!400$ & $2.712 \times 10^8$ & $-1.313(6)$   \cr 
$10$   & $14,\!000$ & $2.764 \times 10^8$ & $-1.919(6)$   \cr 
$11$   & $14,\!300$ & $2.898 \times 10^8$ & $-2.593(7)$   \cr 
$12$   & $14,\!400$ & $2.840 \times 10^8$ & $-3.308(8)$   \cr 
\br
\end{tabular}
\end{indented}
\label{tableEE}  
\end{table}

\begin{table}[h!]
\caption{Estimates of $\gamma_f$ in the BCC lattice}
\begin{indented}
\lineup
\item[]\begin{tabular}{@{}*{5}{c}}
\br                              
$f$& Length & Tours & $\gamma_f$ (BCC)  \cr 
\mr
$3$     & $12,\!000$ & $2.142 \times 10^8$ & $\0\;1.0429(5)$   \cr 
$4$     & $12,\!000$ & $2.106 \times 10^8$ & $\0\;0.8340(8)$   \cr 
$5$     & $12,\!000$ & $3.034 \times 10^8$ & $\0\;0.5405(20)$   \cr 
$6$     & $12,\!000$ & $2.720 \times 10^8$ & $\0\;0.1715(25)$   \cr 
$7$     & $12,\!005$ & $5.262 \times 10^8$ & $-0.259(4)$   \cr 
$8$     & $12,\!000$ & $6.230 \times 10^8$ & $-0.762(6)$   \cr 
\br
\end{tabular}
\end{indented}
\label{tableEF}  
\end{table}

\subsection{Amplitudes}

Estimates of the amplitude $U^{(f)}$ in equation \Ref{14} can be made
by using the results for $\gamma_f$.  Noting that the $o(1)$ term is dominated 
by the confluent correction, it follows that
\begin{equation}
\log(s_n^{(f)}/\mu_d^n) - (\gamma_f {-}1)\log n
 = \log U^{(f)} + B_f \, n^{-\Delta_1} .
\end{equation}
Thus, by taking $\gamma_f$ at its best value, and by plotting the left-hand
side against $n^{-\Delta_1}$, the $y$-intercept would be equal to $\log U^{(f)}$.
To avoid bias due to corrections at small $n$, data with $n\leq 30f$ were discarded,
and the remaining data plotted and extrapolated using a linear model
against $n^{-\Delta_1}$.  This gives the estimates in table \ref{tableCest3}.

The error bars were estimated by exploring the values of the amplitudes
at the limits of the confidence intervals on $\gamma_f$, as shown in tables
\ref{tableE}, \ref{tableEE} and \ref{tableEF}, for the SC, FCC and BCC lattices
respectively.  Notice that the amplitudes $C_k^{(f)}$ in equation
\Ref{2} can be obtained using the symmetry factors relating $U^{(f)}$ 
and $C_k^{(f)}$.  For example, in the cubic lattice the relation is given
by equation \Ref{15} for $3\leq f \leq 6$ and $0\leq k < f$.  This relation similarly
generalises to the FCC and BCC, but for $3\leq f \leq 12$ and $3\leq f \leq 8$,
respectively.

\begin{table}[t!]
\caption{Estimated amplitudes $U^{(f)}$}
\begin{indented}
\lineup
\item[]          
\begin{tabular}{{c}| @{}*{4}{l}}
\br                  
 $f$ &\0 SC lattice & FCC lattice & BCC lattice \cr
\mr
$3$ &\0 $1.142(7)$  &  $1.189(25)$   & $1.161(12)$   \cr
$4$ &\0 $0.952(10)$  &  $1.25(4)$   & $1.147(18)$   \cr
$5$ &\0 $0.631(10)$  &  $1.36(6) $   & $1.09(5)$ \cr
$6$ &\0 $0.247(8)$  &  $1.53(8) $   & $0.95(5)$ \cr
$7$ &\0 $-$              &  $1.58(11)$   & $0.62(5)$ \cr
$8$ &\0 $-$              &  $1.66(15)$   & $0.26(3)$ \cr
$9$ &\0 $-$              &  $1.56(17)$   & $-$ \cr
$10$ &\0 $-$            &  $1.20(13)$   & $-$ \cr
$11$ &\0 $-$            &  $0.80(10)$   & $-$ \cr
$12$ &\0 $-$            &  $0.28(4)$   & $-$ \cr
\br
\end{tabular}
\end{indented}
\label{tableCest3}   
\end{table}

\section{Sampling Branched Structures}
\label{S5}  

In this section the consistency of vertex exponents is examined by 
considering the scaling of more general branched structures.  That is, 
we calculate the entropic exponents of the acyclic branched structures in 
figure \ref{models} to show that they satisfy the relations in 
equation \Ref{15CC} within the numerical accuracy obtained in this paper. 

As in section \ref{section11}, define the \textit{length} or \textit{size} 
$n$ of a lattice network to be the total number of steps or edges.  
A lattice network with connectivity $\mathcal{G}$ is \textit{monodisperse} 
or \textit{strictly uniform} if all branches have the same length. The
underlying connectivities of networks in this study are shown in 
figure \ref{models}, and are $\mathcal{C}$ (a comb) and two kinds 
of brushes, namely $\mathcal{B}_1$ and $\mathcal{B}_2$.  Examples 
of these networks are shown in figures \ref{Comb1}, \ref{Brush1} 
and \ref{Brush2}.  The scaling of $\mathcal{C}$, $\mathcal{B}_1$ 
and $\mathcal{B}_2$ are given by equation \Ref{14CX}, and the 
entropic exponents are given in terms of the vertex exponents in equation
\Ref{15CC}.

There is a repulsion between vertices of higher degree in
networks like combs and brushes which stretches the self-avoiding
walks joining them.  This effect is more difficult to simulate using 
the PERM algorithm, and motived us to use the Wang-Landau algorithm 
\cite{WL01} instead.  In this study we used a parallel implementation 
of this algorithm.

This Wang-Landau
algorithm efficiently approximates the density of states in the presence 
of a cost function. In this case the cost function is the energy of
the states. The sampling is from a probability distribution that becomes 
asymptotically uniform.  If $E_{old}$ ($g(E_{old})$) is the energy 
(respectively density) of the current configuration and $E_{new}$ 
($g(E_{new})$) is the energy (respectively density) of 
the proposed configuration, a proposed move to the new state 
is accepted with probability $\min\LC\frac{g(E_{old})}{g(E_{new})},1\RC$. 
Each time a state is visited, the density of states is updated by 
a modification factor $f$ such that $g(E)\leftarrow g(E)\cdot f$. 
A histogram $H(E)$ recording each visit is kept and a flatness 
criterion for the histogram is used to update the modification
 factor $f$. That is, when the histogram achieves the flatness 
 criterion it is reset and $f$ is reduced in a predetermined fashion. 
 This must be done with care, since if $f$ is decreased too rapidly 
this can lead to saturation errors. In our implementation there 
are 4 parallel streams that are used to control the update of 
$f$ common to all parallel streams.

For branched structures the algorithm first grows a central 
uniform star and then grows the additional branches from
 the endpoint of that star. To grow a star with $f$ arms the 
 central vertex is fixed and at each stage $f$ steps are sampled 
uniformly at random to be appended to the end-vertices of the star. 
If there are no intersections in the proposed steps and 
the state change is accepted then the new configuration is kept. 
Otherwise, the original configuration is re-read and the density 
is updated accordingly. When the star is fully grown the branch 
vertex is chosen uniformly at random from the $f$ candidates. 
Once chosen the remaining branches are grown from the 
branch vertex analogously to the arms of the star. 

Let $b$ denote the number of total branches (including the original 
star arms), each of length $\ell$, of the comb or brush under 
consideration. The process of first growing a star and then 
growing the remaining branches is iterated so that each structure 
of uniform length $n=b\ell$ is independently sampled via the 
Wang-Landau algorithm for $\ell=1,...,200$. For each $\ell$, 
on the order of $10^9$ configurations were sampled.  
A more explicit formulation of the Wang-Landau algorithm 
used for sampling stars is provided below for reference.

\medskip

\noindent \textbf{Wang-Landau Algorithm}

\noindent This algorithm samples $M$ stars with $s$ arms, 
each of length $0$ to $\ell$ and returns the approximate 
counts $\hat c_\ell$ at each length $\ell$. Define $d_\ell=\ln(\hat c_\ell)$

\begin{enumerate}
    \item[1.] Let $d_\ell=0$ for all $\ell$, set $f=1$ and let $v_0$ 
be the vertex at the origin. Set checkpoint $c$ to test for 
histogram flatness and $o_\ell$ the number of observations of 
length $\ell$ for each $\ell$. Let $t=1$ be the number of checks.
    \item[2.] Suppose $\ell>m\geq0$. Choose uniformly among 
the nearest neighbors of $\{v^{1}_m,...,v^{s}_m\}$, 
unoccupied or otherwise, to propose the next steps of the star.
    \item[3.] If the proposed move is $\{v^{1}_{m-1},...,v^{s}_{m-1}\}$ 
then step back with probability $\min\{1,\exp\{d_m-d_{m-1}\}\}$. 
Set $d_{m-1}=d_{m-1}+f$ and $o_{m-1}=o_{m-1}+1$. Otherwise 
reread the current location and set $d_{m}=d_{m}+f$ and 
$o_m=o_m+1$.\\
Else check for intersections with previously visited vertices 
$\{v^{1}_{i},...,v^{s}_{i}\}$ for $i=0,...,m-1$ and amongst 
the proposed vertices. If there are no intersections set 
$\{v^1_{m+1},...,v^{s}_{m+1}\}$ to be the new vertices with 
probability $\min\{1,\exp\{d_m-d_{m+1}\}\}$. Set 
$d_{m+1}=d_{m+1}+f$ and $o_{m+1}=o_{m+1}+1$. Otherwise if 
the proposed vertices are rejected or there is an intersection, 
reread the current location and set $d_{m}=d_{m}+f$ 
and $o_m=o_m+1$.
    \item[4.] Suppose $m=\ell$. Then perform the steps as in 
step 3 but step forward with probability $0$.
    \item[5.] Repeat steps 2 to 4 until $c$ iterations are performed. 
Test for histogram flatness by considering the $o_\ell$. If the 
desired flatness is reached set $t=t+1$ and update $f$.
    \item[6.] Repeat steps 2 to 5 until $M$ observations 
have been reached.
\end{enumerate}

Data were collected and analysed similarly to the analysis done in section 
\ref{S2}.  In determining the approximate counts for lattice networks, 
there are (similarly to the case for lattice stars) symmetry factors which 
should be taken into account when calculating amplitudes $C_{\mathcal{G}}$.

The symmetry factors are determined as follows:  Let the root star of the 
network have $f$ arms, and $b{-}f$ branches are grown on the endpoint 
of one of the root star arms.  The symmetry factor is then equal to 
the number of ways to colour these arms, namely $(f{-}1)!(b{-}f)!$.  
This is seen by noting that the arm from which the branching occurs is
coloured in one way, and the remaining arms in $(f{-}1)!$ ways.  
The last $b{-}f$ arms can be coloured in $(b{-}f)!$ ways. 

In addition, the counts also have to be normalised by counting 
the number of ways the same network can be grown by the 
algorithm.  Each network of length $n$ is grown by first growing 
a star of length $\ell f$ and then growing the addition arms 
comprising $(b{-}f)\ell$ steps.  In $d$ dimensions the sample 
space of each step in the $f$-star is $(2d)^f$ and for the 
additional branches is $(2d)^{b-f}$.  In flat histogram sampling 
these factors are accounted for in the relative weights of 
stars-to-network.  Since stars are grown first and the empty walk 
of unit weight is the root of the star, if the normalization 
is done in this way, there is systematic under-counting by a factor of 
$(2d)^{f-(b-f)}$. Taken together, in order to account for these
factors we must divide the original counts by 
$\frac{(f-1)!(b-f)!}{(2d)^{2f-b}}$.

\subsection{Estimating vertex exponents from lattice networks}
\label{S51}
In this section we will follow closely the procedure for stars in order to 
estimate the entropic exponent of branched lattice networks. These 
estimates can then be used to recover the vertex exponents and 
corroborate the estimates of the preceding sections alongside the 
scaling relation \Ref{14C}. As with stars we expect
\begin{equation}
c_{n}(\C{G}) = C_\C{G}\, n^{\gamma_\C{G} - 1}\, \mu_d^{n}
\end{equation}
and so, to estimate $\gamma_\C{G}$ from the data we search for 
$x=\gamma_\C{G}-1$ such that
\begin{equation}
Q_n(x)=\log\left(\frac{c_{n}(\C{G})}{\mu_d^nn^x}\right)\simeq \hbox{constant}
\end{equation}
as $n\to\infty$. These networks have $b\in\{5,6,7\}$ branches and were 
sampled for $n=b\ell$ where $\ell=1,2,...,200$ (that is, $200$ steps per branch) 
giving a maximum size $n_{max}\in\{1000,1200,1400\}$. To account for 
corrections due to small networks we perform the analysis for 
$\ell\geq\ell_{min}$. By plotting $Q_n(x)$ against $\log(n)$ for 
$\ell\geq\ell_{min}$ and different values of $x$ we can calculate 
the slope by way of a linear fit and interpolate to find the optimal 
value of $x$. Since we have less data for branched structures we 
let $\ell_{min}$ range over $\{2,...,15\}$ and then similarly extrapolate 
the estimate via a fit against 
$\frac{1}{\sqrt{n_{min}}}=\frac{1}{\sqrt{b\ell_{min}}}$.
To arrive at a final estimate for $\gamma_\C{G}$ we first take independent 
samples of $90\%$ of our data to estimate the slope for each $x$. Note that 
for each $\ell$ the counts were generated completely independently so there 
is no serial correlation in the estimates introduced by the simulation. When this 
is done we have several data sets with estimates at each $\ell_{min}$ and we 
subset each data set by sampling half of the $\ell_{min}$ values at random. 
Iterating this several times gives us a large set of estimates from which we can 
calculate the variance. The error bars are given by three standard deviations to 
account for unknown corrections due to the size of the networks and the best 
estimate is taken to be the average. In total we sample the count data $10$ times 
and perform the extrapolation procedure $100$ times. In all, this gives us a final 
set of $1,\!000$ estimates and the final estimates of this analysis are 
\[\gamma_{\mathcal{C}}-1= -0.0731(18), \ \ \gamma_{\mathcal{B}_1}-1
= -0.2896(69), \ \ \gamma_{\mathcal{B}_2}-1= -0.5065(58).  \]
Using the best estimate of $\sigma_1$ in the square and cubic lattices 
allows us to recover the vertex exponents $\sigma_f$ for $f=3$ and 
$f=4$ from the estimates of $\gamma_\C{G}$ via the relations in 
equation \Ref{15CC}.  For $\mathcal{B}_1$ we use the estimate of 
$\sigma_3$ from $\gamma_{\mathcal{C}}$ to get a second estimate 
of $\sigma_4$. The results are compiled in Table \ref{table-expfrombranch} alongside our estimates from stars in the cubic lattice.  The estimates
of $\sigma_4$ slightly overestimate the results from the star data
(in particular the estimate from $\mathcal{B}_2$ which is roughly 
three error bars from the star estimate).  This may be due to the shorter
branches in our network models.  Overall, our network results for
$\sigma_3$ and $\sigma_4$ are in good agreement (for $\mathcal{C}$ 
and $\mathcal{B}_1$), or at worst marginal (for $\mathcal{B}_1$),
when compared to our results from lattice stars.  

Finally, comparing the results for $\gamma_{\mathcal{G}}$ with
the predicted values from our star data via equation \Ref{15CC}
shows very good agreement (see table \ref{table0000}), and 
confirms for these networks the predictions from relation \Ref{0}.

\begin{table}[t]
\caption{Estimates of $\sigma_f$ from $\gamma_\C{G}$ in $3d$.}
\centering
\scalebox{0.875}{        
\begin{tabular}{c | @{}*{2}{c}}
\br                              
Exponent &  Table \ref{table00} & This work \cr 
\mr
$\sigma_3 \ (\mathrm{via} \ \mathcal{C})$ & $-0.19313(11)$ & $-0.19350(91)$ \cr
$\sigma_4 \ (\mathrm{via} \ \mathcal{B}_1)$ & $-0.4802(5)$ & $-0.4885(78) $  \cr
$\sigma_4 \ (\mathrm{via} \ \mathcal{B}_2)$ & $-0.4802(5)$ & $-0.4887(29)$  \cr
\br
\end{tabular}
}
\label{table-expfrombranch}   
\end{table}

\subsection{Estimating amplitudes for lattice networks}
\label{S52}
Taking logarithms of the ratio $c_n(\mathcal{G})/c_n$,
and using equations \Ref{3A} and \Ref{14CX} gives the models
\begin{equation}
\log \LB \Sfrac{c_n(\mathcal{G})}{c_n} \RB 
= \log \LB \Sfrac{C_\mathcal{G}}{C} \RB
+(\gamma_\C{G}-1-2\sigma_1)\log n,
\label{eqn_branchampl1}  
\end{equation}
\begin{equation}
\log \LB \Sfrac{c_n(\mathcal{G})}{\sqrt{c_{2n}}}\RB 
= \log \LB \Sfrac{C_\mathcal{G}}{2^{\sigma_1}\sqrt{C}} \RB 
+ (\gamma_\C{G}-1-\sigma_1)\log n .
\label{eqn_branchampl2}  
\end{equation} 
Fitting the model to our data allows us to get estimates for the amplitude 
ratios $C_{\mathcal{G}}/C$ where $C$ is the self-avoiding walk amplitude
(see equation \Ref{3A}). By considering the results for $\ell_{min}=1,2,...,50$  
we extrapolate using 
\begin{eqnarray}
\log \LB \frac{C_{\mathcal{G}}}{C} \RB \vv_{n_{min}}
&\approx \beta_0 + \frac{\beta_1}{n_{min}} + \frac{\beta_2}{n_{min}^2} , \cr
\log \LB \frac{C_{\mathcal{G}}}{2^{\sigma_1}\sqrt{C}} \RB \vv_{n_{min}}
&\approx \delta_0 + \frac{\delta_1}{n_{min}} + \frac{\delta_2}{n_{min}^2} .
\label{y29}   
\end{eqnarray}
A systematic error is estimated by comparing the results to that of a three parameter 
regression adding the term $c/n$ to the right hand side of equations 
\Ref{eqn_branchampl1} and \Ref{eqn_branchampl2}. Due to data limitations, to 
avoid over fitting in this three parameter fit we perform the extrapolation for $\ell_{min}\in\{1,...,15\}$. The estimated systematic error is taken to be the 
absolute difference between these estimates. Then, by using the best estimate 
of $\sigma_1$ we can solve simultaneously for $C_{\mathcal{G}}$ and $C$. 
Our results are collected in Table \ref{table-constfrombranch}. The final 
reported errors are computed by carrying through the errors computed in 
the original fits. Once again, we see good agreement between the values 
of $C$ for each network structure.

\begin{table}[t]
\caption{Estimates of $C(\mathcal{G})$ in the cubic lattice.}
\centering
\scalebox{0.875}{        
\begin{tabular}{c | @{}*{2}{c}}
\br    
Network &$C_{\mathcal{G}}$ & $C$ \cr 
\mr
$\mathcal{G}=\mathcal{C}$ &\quad $0.29(4)$ & $1.19(6)$ \cr
$\mathcal{G}=\mathcal{B}_1$ &\quad $0.094(5)$ & $1.196(6) $  \cr
$\mathcal{G}=\mathcal{B}_2$ &\quad $0.0304(12)$ & $1.2013(16)$  \cr
\br
\end{tabular}
}
\label{table-constfrombranch}   
\end{table}

\section{Conclusion}
\label{S6}  

In this paper we have given an account of sampling lattice stars and 
acyclic lattice networks using implementations of Parallel PERM 
\cite{G97,HG11,CJvR20} and a parallel implementation of the 
Wang-Landau algorithm \cite{WL01}.   Our simulations produced 
a large set of data in the form of approximate counts, which we 
analysed to extract estimates of the entropic exponents $\gamma_f$, 
and the vertex exponents $\sigma_f$, of lattice stars, as well as 
lattice networks. 

Our final estimates of the entropic exponents are listed in table 
\ref{tafel1} (these are weighted averages of the estimates obtained
in tables \ref{tableE}, \ref{tableEE} and \ref{tableEF}).  Our best estimates
of the vertex exponents are shown in tables \ref{table00} and \ref{table000}
for $3\leq f \leq 6$.  These results are consistent with, and in some cases 
improve on, estimates in other studies. 

The estimates obtained from the $\eps$-expansion are shown in table
\ref{table000}.  For the exponents $\sigma_1$ and $\sigma_3$ 
the $\epsilon$-expansion gives reasonable results at the order $\eps$ 
level, but breaks down at higher orders.  The $\epsilon$ expansions 
for $\sigma_4$, $\sigma_5$ and $\sigma_6$ deteriorate
for higher order expansions, and there appears little prospect at this time,
even with resummation techniques, of finding better values for
the vertex exponents in this way.

Our results for the entropic exponents for the lattice networks 
$\mathcal{C}$, $\mathcal{B}_1$ and $\mathcal{B}_2$ are listed in
table \ref{table0000}.  Our results deviate from predictions of 
first order $\eps$-expansion,  and higher order expansions does not
improve this.   Our predictions of the entropic exponent for lattice
networks from our vertex exponent data are shown in table \ref{table0000},
and compares well with the direct estimates obtained from our
simulations.  This serves as an independent confirmation of the earlier 
results and presents evidence affirming the theorized scaling 
relation \Ref{15CC} for these branched structures.

\vspace{1cm}
\noindent{\bf Acknowledgements:} 
The authors are grateful to P Grassberger for extensive discussions,
sharing of data, and advice while this study was done, and for pointing
out that our estimate of $\sigma_4$ in an earlier preprint was flawed due 
to a biased random number generator.  EJJvR acknowledges financial 
support from NSERC (Canada) in the form of 
Discovery Grant RGPIN-2019-06303.  SC acknowledges the support of 
NSERC (Canada) in the form of an Alexander Graham Bell Canada 
Graduate Scholarship (Application No. CGSD3-535625-2019).

\vspace{2cm}
\noindent{\bf References}
\bibliographystyle{plain}
\bibliography{stars}

\end{document}